\documentstyle[12pt,twocolumn,aps,prb,epsf]{revtex}

\newcommand{\myfig}[3] {
\begin{figure}
\centerline{\epsfysize=#2  \epsfbox{#1.eps}}
\caption{#3}
\label{#1}
\end{figure} }

\def\be{ \begin{equation} }
\def\ee{ \end{equation} }

\begin{document}
\tightenlines
\draft

\title{\bf
Phase Diagram of the three-dimensional Gaussian Random Field Ising Model:
A Monte Carlo Renormalization Group Study
}
\author{M.~Itakura}
\address{JST Domestic Research Fellow,
Center for Promotion of Computational Science and Engineering,
Japan Atomic Energy Research Institute,
Meguro-ku, Nakameguro 2-2-54, Tokyo 153, Japan\\}
\date{\today}
\maketitle

\begin{abstract}
With the help of the replica exchange Monte Carlo method and
the improved Monte Carlo renormalization-group scheme,
we investigate over a wide area in the phase diagram 
of the Gaussian random field Ising model on the simple cubic lattice.
We found that the phase transition at a weak random field
belongs to the same universality class as the
zero-temperature phase transition.
We also present a possible scenario for the
replica symmetry-breaking transition.
\end{abstract}

The random field Ising model in three dimension \cite{review}
has been intensively studied for over 30 years 
by experimental, theoretical, and numerical methods.
But there are still some unanswered questions.
One of the questions is as follows:
Does the phase transition for different strengths of random field
belong to the same universality class?
Critical behavior  at the strong field region \cite{rieger,machta}
and zero-temperature line \cite{zero1}
has been studied numerically by several authors, and 
from the fact that
the observed value of the critical exponent $\theta$ is positive,\cite{rieger}
the existence of the zero-temperature random fixed point (ZRFP)
seems unquestionable.
Together with the fact that 
random field is relevant at the pure critical point, \cite{review}
the renormalization-group (RG)
flow in the temperature-field phase diagram
can be depicted as in Fig. \ref{rg-th}.
The behavior of the RG flow in  the intermediate region
is unknown
and it is unclear whether the ZRFP governs the whole
phase transition or not.

Another question is as follows:
Is there a glasslike phase near the critical
line in the phase diagram?
Theoretical studies based on the
replica formalism \cite{review} have predicted the
presence of a glasslike phase which is
characterized by replica symmetry breaking
near the transition temperature. \cite{review}
But owing to the unphysical $n\rightarrow 0$ limit 
in the theory, interpretation of the result is not trivial.

For the investigation of these problems,
direct observation of the RG flow may be the most powerful
method.
In the present work, we numerically observed
the RG flow of the Gaussian random field Ising model
over a wide region in the phase diagram
using an improved Monte Carlo renormalization-group (MCRG) scheme, \cite{mcrg}
which is a very simple and sophisticated method.
%and found that
%the phase transition at any field strength
%belongs to the same universality as the 
%ZRFP.
%We also show that 
%in the paramagnetic phase
%there are crossover between 
%glass-like behavior and non-glass-like behavior,
%but no phase transition between them.  
%%%%MODEL
We used the following model in the Monte Carlo simulations:
\be
H=\sum_{<ij>} S_i S_j +h \sum_i S_i h_i,
\ee
where the first summation runs over all the nearest-neighbor
pairs of the sites on an $L\times L \times L$ simple cubic lattice
(up to $L=16$)
with a periodic boundary condition, and $h_i$ is a random
Gaussian variable whose
average is $0$ and variance is $1$.
%%%%%OBS
We measured the following quantities: 
\begin{eqnarray}
T_L &=&{ [ \langle (M^2-\langle M^2\rangle )^2\rangle ] \over 2 [\langle M^2\rangle ]^2},  \\
%   &=&{ [<M^4>] \over 2 [<M^2>]^2}-{[<M^2>^2]\over 2 [<M^2>]^2},\\
S_L &=&{ [ (\langle M^2\rangle -[\langle M^2\rangle ])^2]\over 2 [\langle M^2\rangle ]^2}, 
%    &=&{ [<M^2>^2] \over 2 [<M^2>]^2} -1, 
\end{eqnarray}
where $M$ denotes magnetization,
the symbol $\langle \cdots \rangle $ and $[\cdots ]$ denote 
thermal and ensemble averages, respectively.
$T_L$ and $S_L$ are the amplitudes of thermal and
 sample-to-sample
fluctuations, respectively, of the rescaled 
block spin $M / \sqrt{[\langle M^2\rangle ]}$ 
defined on an $L\times L\times L$
block. \cite{mcrg}
Thus $T_L$ and $S_L$ reflects the renormalized temperature and
strength of the random field, respectively.
%%%%%RG
If the ZRFP governs the whole critical line,
the flow of $T_L$ and $S_L$ will become the one depicted in
Fig. \ref{rgth}(a).
On the other hand, if there exists another fixed point,
the flow will become the one depicted in Fig. \ref{rgth}(b).
Note that, in the paramagnetic phase,
thermal and  sample-to-sample fluctuations 
are on the same order $O(L^{3/2})$
and there are a set of ``high temperature fixed points''
on a line $T_L+S_L=1$, where both 
fluctuations are Gaussian.
The position of the fixed point
depends on the ratio between the two amplitudes.

%%%%%%%%%%MC 
In the Monte Carlo simulation,
we used a single-spin-flip Metropolis update and
the replica exchange method. \cite{repex}
%More elaborated method such as cluster
%exchange method \cite{clustex} does not always
%accelerate the simulation
%owing to the heavy computational cost.
%But the additional computational cost for 
%replica exchange is negligible and
%the replica exchange accelerated the simulation
%drasicaly, as we mention later.
In a replica exchange procedure,
an exchange for all adjacent temperature pairs
was tried. The exchange procedure was performed
for every two Metropolis sweeps.
We used 20 replicas for all field realizations.
Temperatures for all replicas were chosen so that
the replica exchange occurs with modest frequency and
the highest temperature is well above the transition
temperature where the single-spin-flip provides good ergodicity.
Figure \ref{gph} shows the region in the phase diagram
where the simulations were carried out in the present work,
together with other recent numerical studies.
The case $h=1.3$ was the hardest to relax,
in which we used 88 000 Monte Carlo steps
(discarding the initial 8000 steps)
for each field realization. 
We checked that two kinds of initial states,
all spins up and down, give the same thermal averages
of $M$ within statistical errors, thus the system is
well equilibrated.
This result is remarkably fast compared to 
the simulation in which only the Metropolis update 
was used and $2 \times 10^6$ steps were needed to fulfill the
same criterion. \cite{rieger}
We used 120 different random field realizations for each $L$ and
$h$. This number of samples is  rather small compared to
other recent numerical works. But our aim in the present
work is to draw
qualitative conclusions, not quantitative estimates of critical
exponents, 
and it was enough for this purpose.

%%%%RESULT
Figure \ref{mcrg} shows the flow of $T_L$ and $S_L$ on a logarithmic scale.
All lines are drawn from 
$(T_L,S_L)$ to $(T_{2L},S_{2L})$ with $L=4$ (thin lines) and $L=8$ (bold lines).
The RG flow of both sizes are consistent, which indicates that
the finite-size effect is small enough and the final destination of the RG flow
can be  safely predicted.
Thus one can see that
there are no fixed points in the
intermediate region, and
the renormalization flow which starts 
at the weak-field region is eventually attracted to
the low-temperature region $T_L < 0.01$.
%%%%%RGFLOW UNIV
Figure \ref{rmcrg} shows the RG flow near the ZRFP.
The precise position of the ZRFP $(0,S_*)$ is hard to
estimate, owing to its proximity to the low-temperature
fixed point $(0,0)$. \cite{note0} 
We only performed a rough estimate of
the critical value $S_*$ as $0\leq S_* \leq 0.01$,
which is consistent with the result of the zero-temperature
simulation, \cite{zero1} $S_* = 0.003(2)$.
%Note that
%$<M^n> = M_0^n$ ($M_0$ denotes ground state magnetization)
%and $S_L = [M_0^4]/2[M_0^2]^2 -1$ at zero-temperature.
Thus Figs. \ref{mcrg} and \ref{rmcrg}
indicate that the ZRFP governs the whole
phase-transition line.

%%%FIRST ORDER
Now let us consider the possibility of a first-order transition.
If the transition is of first order,
an ensemble distribution of $\langle M^2\rangle $ will exhibit a
double-peak
structure and a plot of $S_L$ will develop a sharp peak
at the transition temperature.
But no such behavior was observed in the present work.
%Of course there always remains a possibility that
%the first order behavior emerges in the larger system sizes
%and one can never exclude that possibility.
Note that when we observe a single sample at zero temperature
and varying $h$, the system jumps between different 
ground states and there are many discontinuous transition
points on the $h$ axis. Similar behavior will be observed
in the low-temperature region and somes sample may exhibit
the double-peak behavior. \cite{machta}
However, if ensemble distribution of
the ground-state magnetization at zero temperature
does not exhibit double-peak behavior,
an ensemble average of the ground state magnetization
will not change discontinuously and the transition is of
second order, as is the result of Ref.\cite{zero1}.

%%%%%RSB
Figure \ref{sl-k} shows plots of $S_L$ against temperature
at $h=1.3$. The bold line denotes
the expected behavior in the $L\rightarrow \infty$ limit,
assuming that the basin of the fixed point $S_L=1, T_L=0$ has a
finite measure in Fig. \ref{rgth}(a).
$S_\infty$ reaches its maximum value $1$ slightly above
the critical temperature and remains $1$, then
discontinuously drops to a critical value $S_*$ at the
critical temperature.
Note that $S_L$ is a similar quantity to the replica interaction
term $\phi_\alpha ^2 \phi_\beta^2$ in the
$\phi^4$ model based on the replica formalism, \cite{note1}
in which divergence of the coefficient
of $\phi_\alpha ^2 \phi_\beta^2$ at the paramagnetic phase 
is reported. \cite{brezin}
In the $S_\infty =1$ region,
thermal fluctuation of the magnetization becomes
more and more negligible compared to its absolute value as
$L$ becomes large.
This means that the breadth of each valley in the free-energy
landscape becomes negligible compared to the distances between
each valley.
In this limit, an infinitesimal change of $h$ or $T$ will
drive the system out of a valley and into another valley. \cite{dg}
This could be a possible scenario for the 
replica symmetry-breaking transition \cite{review,brezin}
and can be tested numerically by checking the following
inequality:
\begin{eqnarray}
\lim_{\Delta\rightarrow 0}\lim_{L \rightarrow \infty}
[\langle M(h)\rangle \langle M(h+\Delta)\rangle ] \nonumber \\
\neq [\langle M(h)\rangle ^2],
\label{rsbneq}
\end{eqnarray}
with a very careful finite-size scaling analysis.

%%%%%CONC
In conclusion,
we performed a replica exchange Monte Carlo simulation
of the Gaussian random field Ising model in three dimensions
and, using an improved MCRG scheme, we found that
the phase transition at the weak random field belongs to
the same universality as the zero-temperature phase transition,
and we presented a possible scenario
for the replica symmetry-breaking transition.
\\

The present author thanks K.Hukushima for his useful comments.

%%%%%%%%%%%%%%%%%%%%%%%%%%%%%%%%%%%
\myfig{rg-th}{6cm}{Schematic renormalization flow
in the temperature-field phase diagram.}

\myfig{rgth}{4cm}{Expected renormalization flow
of $T_L$ and $S_L$ for the case of
(a) the zero-temperature random fixed point only
and (b) another fixed point in the intermediate region.
L, P, and $R_x$
denote low-temperature, pure-critical, and random fixed points,
respectively.
}

\myfig{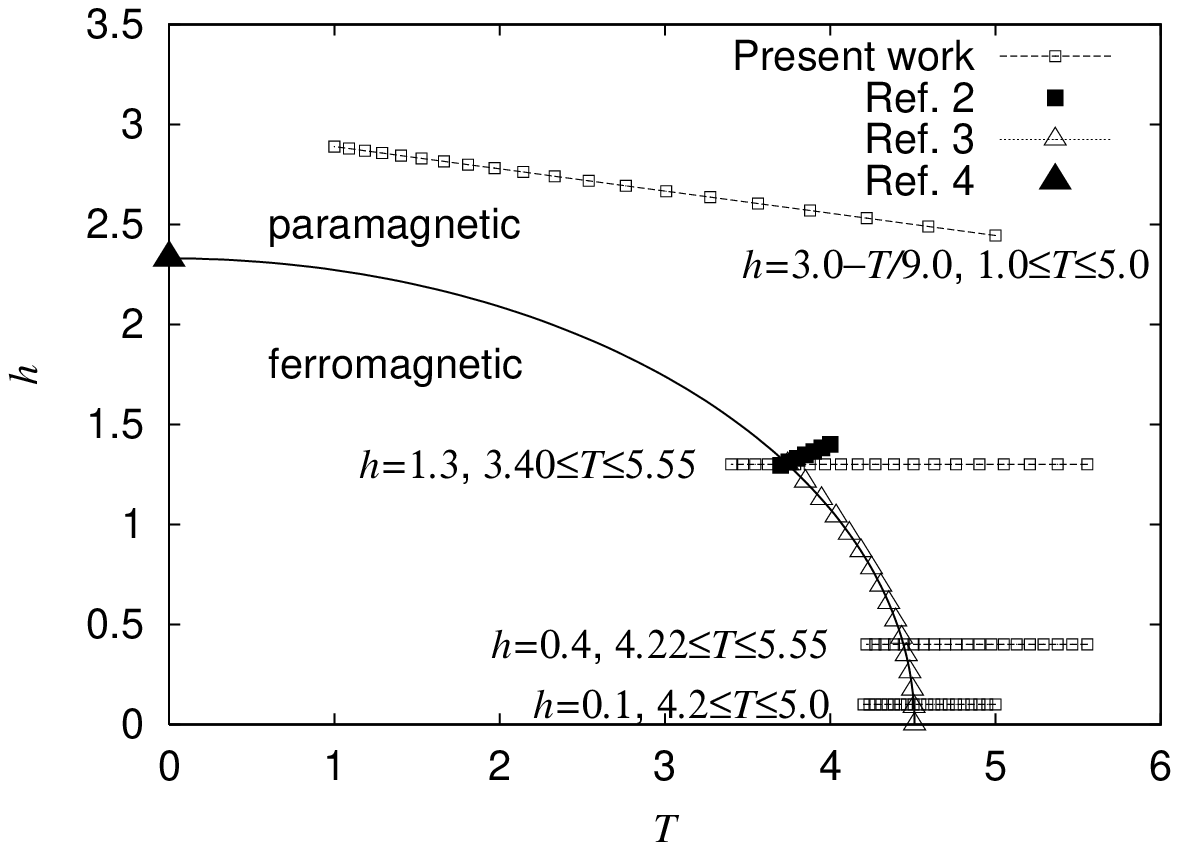}{6cm}{Regions in the $T$-$h$ phase diagram
where simulations were carried out in the present
work, together with other recent numerical works.}

\myfig{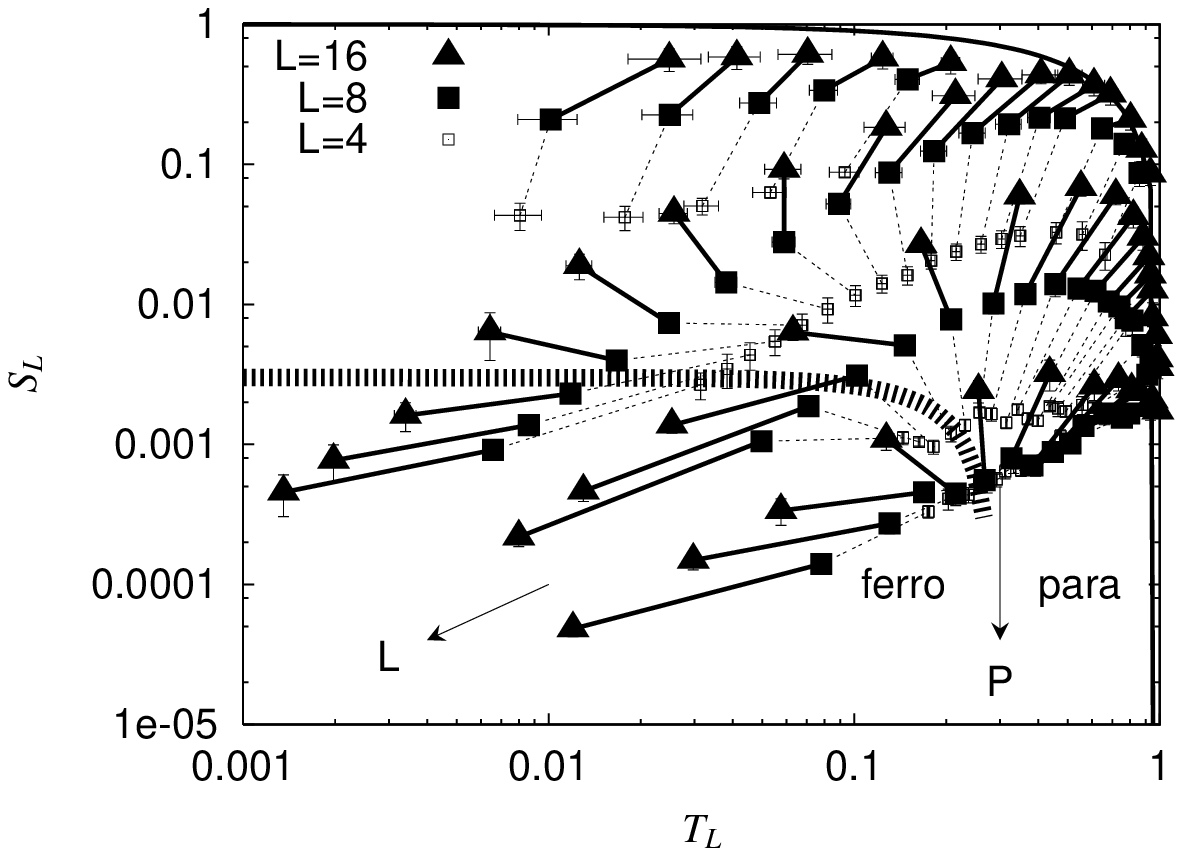}{6cm}{Observed renormalization flow
of $T_L$ and $S_L$. 
The bold dashed line is an approximate phase boundary.
Note the logarithmic scale of the plot. 
Errors are smaller than the size of symbols,
unless errorbars are explicitly shown.
}

\myfig{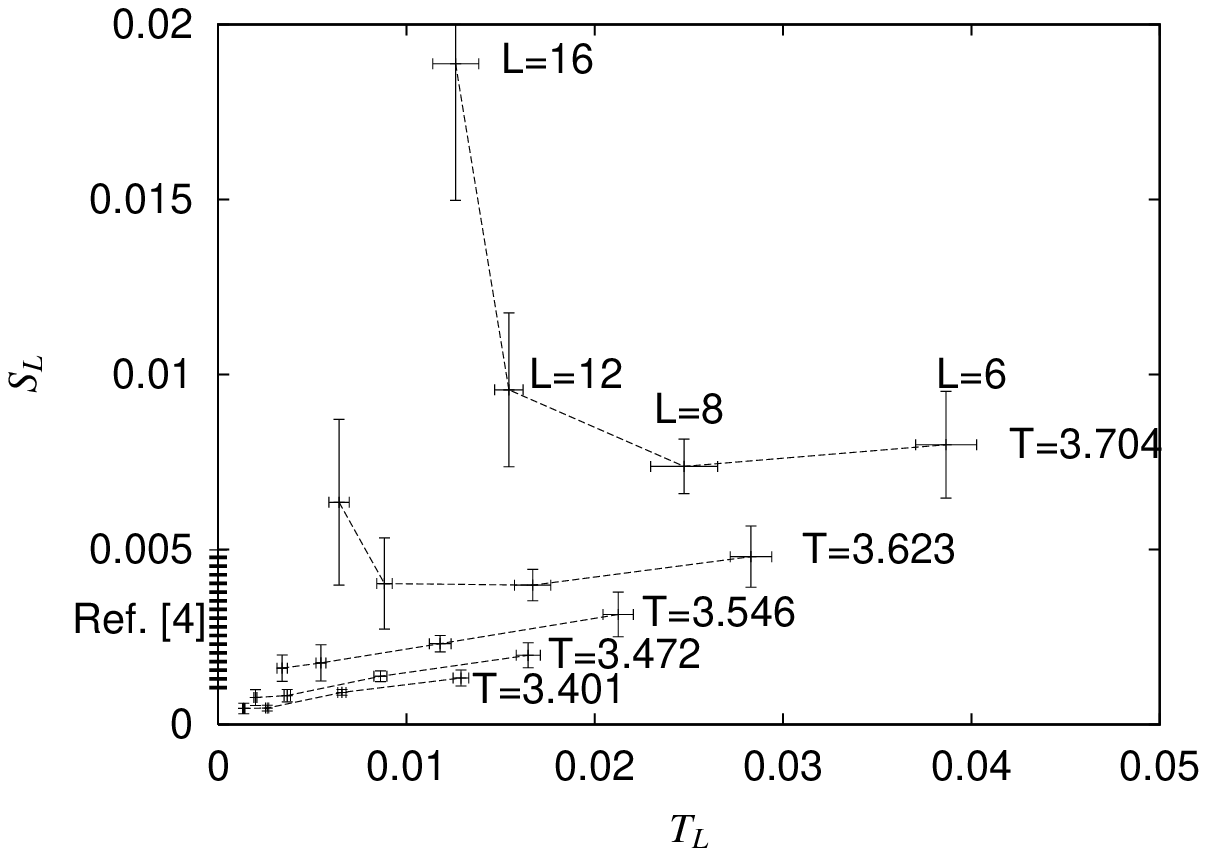}{6cm}{Renormalization flow
of $T_L$ and $S_L$ near the ZRFP.
The bold dashed line is an estimated position
of ZRFT in Ref.[4].
}

\myfig{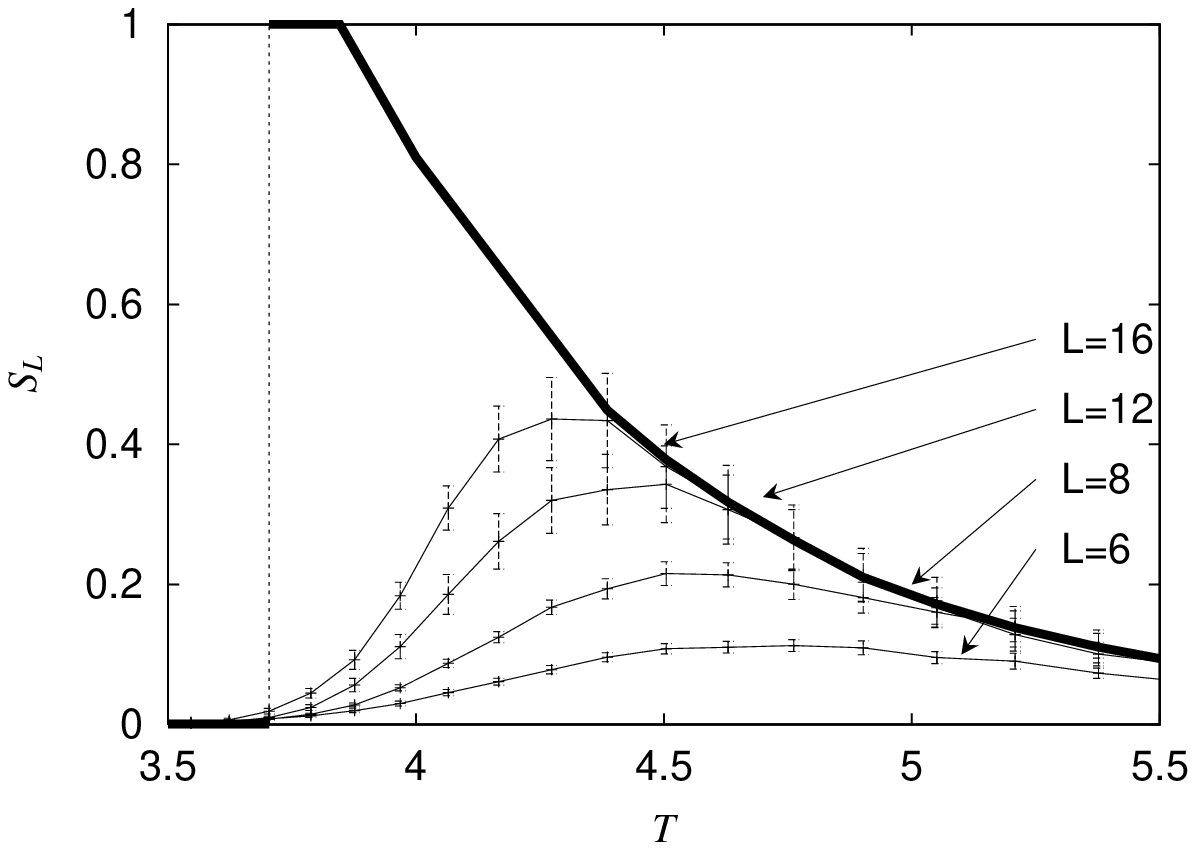}{6cm}{Plots of $S_L$ against temperature $T$
for $h=1.3$. The bold line is the expected behavior for 
the $L=\infty$
case.}


\begin{thebibliography}{99}
\bibitem{note1}
When $\alpha\neq\beta$,
$[\langle \phi_\alpha^2 \phi_\beta^2\rangle ]=[\langle \phi_\alpha^2\rangle ^2]$ for finite systems,
while $S_L=[\langle \phi^2\rangle ^2]/2-1$ if we define $\phi\equiv M/\sqrt{[\langle M^2\rangle ]}$.
\end{thebibliography}
\end{document}